\documentclass{ws-procs975x65}
\usepackage{graphicx}
\usepackage{epstopdf}
\usepackage{latexsym}
\usepackage{amsmath}

\begin{document}

\title{RENORMALIZATION GROUP ANALYSIS OF QUANTUM CRITICAL POINTS
IN $d$-WAVE SUPERCONDUCTORS}

\author{MATTHIAS VOJTA\cite{addr}, YING ZHANG, AND SUBIR SACHDEV}

\address{Department of Physics, Yale University\\
P.O. Box 208120, New Haven, CT 06520-8120, USA}


\begin{abstract}
We describe a search for
renormalization group fixed points which control a
second-order quantum phase
transition between a $d_{x^2-y^2}$ superconductor
and some other superconducting ground state.
Only a few candidate fixed points are found.
In the finite temperature ($T$) quantum-critical region of
some of these fixed
points, the fermion quasiparticle lifetime is very short and
the spectral function has an energy width of order $k_B T$ near
the Fermi points. Under the same conditions,
the thermal conductivity is infinite in the
scaling limit. We thus provide simple, explicit, examples of quantum
theories in two dimensions for which a purely
fermionic quasiparticle description
of transport is badly violated.
\end{abstract}
 \bodymatter

%%%%%%%%%%%%%%%%%%%%%%%%%%%%%%%%%%%%%%%%%%%%%%%%%%%%%%%%%%%%%%

\section{Introduction}
\label{intro}

The quasiparticle excitations of the $d_{x^2-y^2}$-wave high
temperature superconductors have been subjected to intense
scrutiny in the past few years. An especially important test of
our understanding of the underlying physics is whether
quasiparticle relaxation processes measured by different
experimental probes can be reconciled with each other. A striking
dichotomy appears to have emerged recently in such a context.
Photoemission experiments~\cite{valla} indicate that the nodal
quasiparticles have very short lifetimes in the superconducting
state, with their spectral
functions having linewidths of order $k_B T$. In contrast,
transport experiments, especially measurements of the thermal
conductivity, when interpreted in a simple quasiparticle model,
indicate far larger quasiparticle lifetimes in the
superconductor~\cite{ong}.

\begin{figure}[t]
\centering \includegraphics[width=3in]{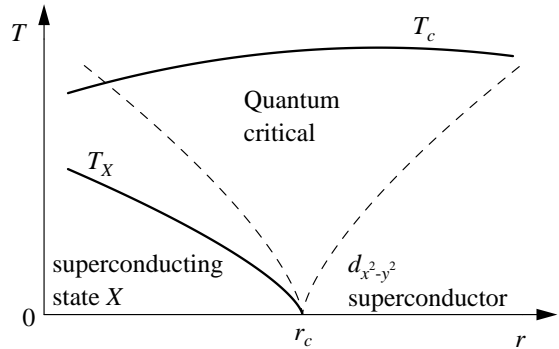} 
\caption{
Finite-temperature phase diagram \protect\cite{vzs}
of the $d$-wave superconductor close to a quantum-critical
point.
Superconductivity is present for $T<T_c$.
The long-range order associated with the state
$X$ vanishes for $T>T_X$, but fluctuations of this order provide
anomalous damping of the nodal quasiparticles in the
quantum-critical region. The tuning parameter $r$ is some coupling
constant in the Hamiltonian, with the suitable choice depending
upon the identity of the state $X$. It is possible, although not necessary,
that increasing
$r$ corresponds to increasing doping concentration, $\delta$. The
critical point $r=r_c$ is not required to be in the experimentally
accessible parameter regime, just not too far from the $d_{x^2-y^2}$
superconductor.
}
\label{fig1}
\end{figure}

Motivated primarily by the photoemission experiments~\cite{valla},
we recently proposed a scenario~\cite{vzs,vzs2} in which the large
linewidths in the fermion spectral functions are explained by
fluctuations near a quantum critical point between the
$d_{x^2-y^2}$ superconductor and some other superconducting state
$X$ (see Fig~\ref{fig1}). In this paper we will review and extend
earlier arguments which classify various possibilities for the
state $X$.
We will provide the details of a renormalization group
(RG) analysis which shows that only a small number of the
candidates for $X$ are associated with a RG fixed point which
describes a second-order phase boundary between $X$ and
the $d_{x^2-y^2}$ superconductor that can be
generically crossed by varying a
single parameter $r$; we will identify the subset of these
fixed points which lead to fermion spectral linewidths of
order $k_B T$. We will also initiate a discussion of the transport
properties of these fixed points, and argue that they offer
attractive possibilities for explaining the transport experiments
as well.

We begin with a brief review of studies of quantum phase
transitions in the cuprate superconductors. The subject was
initiated in the work of Chakravarty {\em et al.\/}\cite{CHN},
who presented a
field-theoretic study of the destruction of N\'{e}el order in
insulating antiferromagnets, but focused mainly on thermal
fluctuations above a N\'{e}el ordered state.
Subsequent work examined the nature of the
paramagnetic ground state in the insulator~\cite{rs}, and
of its quantum-critical point to the N\'{e}el state~\cite{CSY,science}.
In particular, it was proposed~\cite{CSY} that destroying the N\'{e}el order
by adding a finite density of mobile charge carriers also led a
quantum-critical point in the same universality class as in an
insulating antiferromagnet, with dynamic critical exponent $z=1$.
This scenario had strong consequences for magnetic experiments,
and for the manner in which a spin `pseudo-gap' would
develop as $T$ was lowered, and
these appear to be consistent with observations:
crossovers in NMR relaxation rates~\cite{nmr}, uniform
susceptibility~\cite{CSY}, and dynamic neutron scattering~\cite{gabe}
all indicate that
$z=1$. Moreover, a further consequence~\cite{CSY,pwa} of such a scenario was that
the paramagnetic state should have a stable $S=1$ `resonant' spin
excitation near the antiferromagnetic wavevector, and this is also
borne out by numerous neutron scattering studies~\cite{resonance}.

If we accept that the mobile charge carriers have a
superconducting ground state (in particular, a $d$-wave superconductor),
then the arguments for the common
universality of the magnetic quantum critical point in insulating
and doped antiferromagnets can be sharpened.
For both cases, it is clear that the order parameter is a
3-component real field, $N_{\alpha}$, which measures the amplitude
of the local antiferromagnetic order. In the
paramagnetic state, $N_{\alpha}$ will fluctuate about $N_{\alpha}
= 0$, and indeed, the triplet `resonance' modes just mentioned are
the 3 normal mode oscillations of $N_{\alpha}$. The transition
to the state with magnetic long-range order (we assume that
the charge sector is superconducting on both sides
of the transition) is described by the condensation of
$N_{\alpha}$, and the theory for the quantum-critical point will
depend upon whether the $N_{\alpha}$ couple efficiently to other
low-energy excitations, not directly associated with the magnetic
transition. In a $d$-wave superconductor, the important candidates for
these low energy excitations are the fermionic $S=1/2$ Bogoliubov
quasiparticles (one can also consider fluctuations in the phase of
the superconducting order parameter, but we will argue below that
such a coupling can be safely neglected).
Momentum conservation now plays a key role: fluctuations of
$N_{\alpha}$ occur primarily at a finite wavevector ${\bf Q}$
(in the present situation, this is the wavevector at which the
magnetic order appears), and the fermions will
be scattered by this the wavevector. If ${\bf Q}$ does not equal
the separation between two nodal points of the $d$-wave
superconductor [the nodes are the locations in the Brillouin zone
of gapless fermionic excitations, and they are
at $(\pm K, \pm K)$ with $K = 0.39 \pi$
at optimal doping], then it is not difficult to show that the
fermion scattering
serves mainly to renormalize the parameters in the effective low
energy action for the $N_{\alpha}$, and does not lead to any
disruptive low energy damping~\cite{vs}. In such a situation, there is no
fundamental difference between the magnetic fluctuations in a
superconductor and an insulating paramagnet, and both cases have the same
theory for the quantum critical point to the onset of long-range
magnetic order. Conversely, if ${\bf Q} = (2K,2K)$, the coupling to the fermionic
quasiparticles is important, and a new theory obtains: such a
theory was discussed by Balents {\em et al.\/}\cite{bfn}

%It is the fermionic quasiparticles which are the focus of interest
%in this paper, not the magnetic excitations.
In this paper we will focus on the fermionic quasiparticles rather
than the magnetic excitations.
We are interested in damping
mechanisms for the fermions and associated possibilities for the
state $X$ in Fig~\ref{fig1}. The above discussion on
magnetic properties suggests a natural possibility for $X$: a
state with co-existing magnetic and superconducting order.
However, it should also be clear from the discussion above that
strong damping of the fermionic quasiparticles requires ${\bf Q} =
(2K, 2K)$. For the current experimental values, this
condition is far from being satisfied, and so the magnetic
ordering transition is just as in an insulator.
This also means that the magnetic quantum
critical point is not currently a favorable candidate for the quantum
critical point in Fig~\ref{fig1}. We are therefore led to a search
for other possibilities, and this paper will describe the results
of such a search. The new cases we will consider do not have
magnetic order parameters, and so we are envisaging two distinct
quantum critical points near the $d$-wave superconductor: one
involving the magnetic order parameter which is already known to
occur with decreasing doping (and which, as discussed above, is in accord with
numerous magnetic experiments), and another one associated
with the state $X$ which may or may not be present along the
experimentally accessible parameter regime.
For completeness, we will also discuss the
magnetic case with ${\bf Q} = (2K, 2K)$ in Section~\ref{afm}:
this is the only case which envisages a single quantum phase transition to
explain both the magnetic experiments and the fermion damping.

In Section~\ref{rg}, we present the details of a
renormalization group (RG) analysis which searches for candidate fixed
points describing the quantum phase transition between
$X$ and the $d$-wave superconductor: among the many a
priori possibilities, only a few stable fixed points are found.
The structure of the $T \geq 0$ single fermion Green's functions near these
fixed points is discussed in Section~\ref{single}: these results
can be compared with photoemission experiments.
Spin, thermal, and charge transport
properties of all the fixed points are subsequently considered in
Section~\ref{trans}; a significant result is that
the thermal conductivity of the quantum field theories describing the
vicinities of these fixed points is
infinite in the scaling limit.

\section{Renormalization group analysis}
\label{rg}

Let us assume that the order parameter associated with
$X$ carries total momentum ${\bf Q}$. We have argued
\cite{vzs,vzs2} above that its coupling to the fermions can be relevant
only if a nesting condition is satisfied: order parameter
fluctuations will scatter fermions by a momentum ${\bf Q}$, and
such scattering events are important only if they occur between
low energy fermions, {\em i.e.}, the wavevector ${\bf Q}$ (or an
integer multiple of ${\bf Q}$) must equal the separation between
two nodal points. If such a condition is not satisfied, then, as
noted above,
fermion scattering events can be treated as virtual processes
which modify the coupling constants in the effective action, but
do not lead to a fundamental change in the form of the low energy
theory.

Three natural possibilities can satisfy the
nesting condition~\cite{caveat}: ${\bf Q}$ =0, ${\bf Q} =
(2K,2K)$, and ${\bf Q}=(2K, 0),(0,2K)$. We will consider these
possibilities in turn in the following subsections. Of these, the
first condition can be naturally satisfied for a range of
parameter values, while the last two require fine-tuning
unless, for some reason, there is a mode-locking between the values
of ${\bf Q}$ and $K$.

\subsection{Order parameters with ${\bf Q} =0$}
\label{q0}

We will assume that the ${\bf Q}=0$ order parameter is a
spin-singlet fermion bilinear (spin triplet condensation at ${\bf
Q}=0$ would imply ferromagnetic correlations which are unlikely to be
present, while order parameters involving higher-order fermion
correlations are not expected to have a relevant coupling to the
fermions~\cite{caveat}). Simple group theoretic arguments
\cite{vzs2} permit a complete classification of such order
parameters. The order parameter for $X$ must be built out of
the following correlators ($c_{{\bf q} a}$ annihilates an electron
with momentum ${\bf q}$ and spin $a=\uparrow , \downarrow$)
\begin{eqnarray}
\langle c_{{\bf q} a}^{\dagger}
c_{{\bf q} a} \rangle &=& A_{\bf q} \nonumber \\
 \langle c_{{\bf q} \uparrow} c_{-{\bf q}
\downarrow} \rangle &=& \left[ \Delta_0 (\cos q_x - \cos q_y) +
B_{\bf q} \right] e^{i \varphi},
\label{orders}
\end{eqnarray}
where $\Delta_0$ is the background $d_{x^2-y^2}$ pairing which is
assumed to be non-zero on both sides of the transition,
$\varphi$ is the overall phase of the superconducting order,
and $A_{\bf q}$ and $B_{\bf q}$ contain the possible order
parameters for the state $X$ corresponding to condensation in the
particle-hole (or
excitonic) channel or additional particle-particle pairing
respectively.
It is clear that $\varphi$ has the usual charge 2 transformation
under the electromagnetic gauge transformation, and so gradients
of $\varphi$ measure flow of physical electrical current; because
of the non-zero superfluid density associated with $\Delta_0$,
$\varphi$ fluctuations remain non-critical and we will show that
they can be neglected.
It follows that the order parameter $B_{\bf q}$,
which is in general a complex number, carries no electrical
charge. Similarly, $A_{\bf q}$ is also neutral but must be real.

\begin{figure}[t]
\centering \includegraphics[width=2.5in]{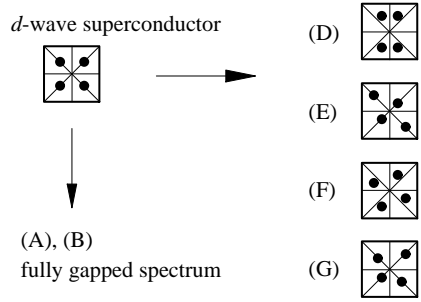} 
\caption{
Evolution of the nodal points when passing from $r>r_c$ to
$r<r_c$ at $T=0$ for 6 of the ${\bf Q}=0$ states $X$ discussed in
Section~\protect\ref{q0}. There is no change in the position
of the nodal points for case C, only a change in velocity.
}
\label{fig1a}
\end{figure}

To classify the order parameters for $X$, we expand $A_{\bf q}$
and $B_{\bf q}$ in terms of the basis functions of the irreducible
representation of the tetragonal point group $C_{4v}$.
This group has 4 one-dimensional representations, which we label
as (basis functions in parentheses) $s$ (1), $d_{x^2-y^2}$ ($\cos
q_x - \cos q_y$), $d_{xy}$ ($\sin q_x \sin q_y$), and $g$
($\sin q_x \sin q_y(\cos
q_x - \cos q_y)$), and one 2-dimensional representation $p$ ($\sin
q_x, \sin q_y$). An analysis of all excitonic order and additional
pairings (both real and imaginary) in these functions has been carried
out~\cite{vzs2}, and it was found that 7 inequivalent order
parameters are allowed for $X$.
Of these, the first 6 (A-F) involve a one-dimensional
representation of $C_{4v}$, and so the order parameter is
Ising-like and represented by a real field $\phi$, while the 7th
(G) involves the 2-dimensional representation and 2 real fields
$\phi_x, \phi_y$. The 7 possibilities for $X$ are
\begin{eqnarray}
&& \mbox{(A) $is$ pairing: $A_{\bf q} = 0$, $B_{\bf q} = i \phi$}
\nonumber \\
&& \mbox{(B) $id_{xy}$ pairing: $A_{\bf q} = 0$, $B_{\bf q} = i \phi \sin q_x \sin q_y $}
\nonumber \\
&& \mbox{(C) $ig$ pairing: $A_{\bf q} = 0$, $B_{\bf q} = i \phi \sin q_x \sin q_y
(\cos q_x - \cos q_y)$}
\nonumber \\
&& \mbox{(D) $s$ pairing: $A_{\bf q} = 0$, $B_{\bf q} =  \phi$}
\nonumber \\
&& \mbox{(E) $d_{xy}$ excitons: $A_{\bf q} = \phi \sin q_x \sin q_y $, $B_{\bf q} = 0$}
\nonumber \\
&& \mbox{(F) $d_{xy}$ pairing: $A_{\bf q} = 0$, $B_{\bf q} =  \phi \sin q_x \sin q_y $}
\nonumber \\
&& \mbox{(G) $p$ excitons: $A_{\bf q} = \phi_x (\sin q_x + \sin q_y)
+\phi_y (\sin q_x - \sin q_y) $, $B_{\bf q} = 0 $}
\label{ordersa}
\end{eqnarray}
Apart from case C, these possible orderings for $X$ change the
nature of the nodal excitations, and these are sketched in
Fig~\ref{fig1a}.
Cases A, B open up a gap in the fermion spectrum over the entire
Brillouin zone (suggesting an energetic preference for these
cases), case C leaves the positions of the nodal points unchanged
but changes a velocity in the dispersion relation, while
cases D--G break $C_{4v}$ symmetries by moving the nodal points as
shown.

Knowledge of the order parameters in (\ref{ordersa}) and simple
symmetry considerations allow us to write down the quantum field
theories for the transition between the $d_{x^2-y^2}$
superconductor and $X$.

The first term in the action, $S_{\Psi}$, is simply
that for the low energy fermionic excitations in the $d_{x^2-y^2}$
superconductor.
We denote the
components of $c_{{\bf q}a}$ in the vicinity of the four nodal
points $(\pm K, \pm K)$
by (anti-clockwise) $f_{1a}$, $f_{2a}$, $f_{3a}$, $f_{4a}$,
and introduce the 4-component Nambu spinors $\Psi_{1a} =
(f_{1a}, \varepsilon_{ab} f_{3b}^{\dagger})$
and  $\Psi_{2a} =
(f_{2a}, \varepsilon_{ab} f_{4b}^{\dagger})$ where
$\varepsilon_{ab}=-\varepsilon_{ba}$ and
$\varepsilon_{\uparrow \downarrow} = 1$ [we will follow the
convention of writing out spin indices ($a,b$) explicitly, while indices
in Nambu space will be implicit].
Expanding to linear order in gradients from the nodal points,
we obtain
\begin{eqnarray}
S_{\Psi} &=& \int \!\! \frac{d^2 k}{(2 \pi)^2} T \!\sum_{\omega_n}
\Psi_{1a}^{\dagger}  \left(
- i \omega_n + v_F k_x \tau^z + v_{\Delta} k_y \tau^x \right) \Psi_{1a}  \nonumber \\
&+& \int \!\! \frac{d^2 k}{(2 \pi)^2}
T \! \sum_{\omega_n}
\Psi_{2a}^{\dagger}  \left(
- i \omega_n + v_F k_y \tau^z + v_{\Delta} k_x \tau^x \right) \Psi_{2a} .
\label{dsid1}
\end{eqnarray}
Here $\omega_n$ is a Matsubara frequency,
$\tau^{\alpha}$ are Pauli matrices which act in the fermionic
particle-hole space, $k_{x,y}$ measure the wavevector from the nodal points and
have been rotated
by 45 degrees from $q_{x,y}$ co-ordinates,
and $v_{F}$, $v_{\Delta}$ are velocities.

The second term, $S_{\phi}$ describes the effective action for the
order parameter alone, generated by integrating out high energy
fermionic degrees of freedom. For cases A--F this is the familiar
field theory of an Ising model in 2+1 dimensions
\begin{equation}
S_{\phi} = \int \!\! d^2 x d \tau \Big[
\frac{1}{2}(\partial_{\tau} \phi)^2 + \frac{c^2}{2} (\nabla \phi )^2 +
\frac{r}{2} \phi^2 + \frac{u_0}{24} \phi^4 \Big];
\label{dsid3}
\end{equation}
here $\tau$ is imaginary time,
$c$ is a velocity, $r$ tunes the system across the
quantum critical point, and $u_0$ is a quartic self-interaction.
For case G, the generalization of $S_{\phi}$ is
\begin{eqnarray}
\widetilde{S}_{\phi} &=& \int \!\! d^2 x d \tau \left[\frac{1}{2}
\left\{
(\partial_{\tau} \phi_x)^2 + (\partial_{\tau} \phi_y)^2 +
c_1^2 (\partial_x \phi_x )^2 +
c_2^2 (\partial_y \phi_x )^2 + c_2^2 (\partial_x \phi_y )^2 \right. \right.
\label{p1} \\
&+& \left.\left. c_1^2 (\partial_y \phi_y )^2 +
e (\partial_x \phi_x ) (\partial_y \phi_y)
+  r (\phi_x^2 + \phi_y^2) \right\}  + \frac{1}{24}
\left\{u_0 (\phi_x^4 + \phi_y^4) + 2 v_0 \phi_x^2 \phi_y^2 \right\}
\right].
\nonumber
\end{eqnarray}

The final term in the action, $S_{\Psi\phi}$ couples the bosonic and fermionic
degrees of freedom. From (\ref{ordersa}) we deduce for A--F
that
\begin{equation}
S_{\Psi\phi} = \int \!\! d^2 x d \tau \Big[ \lambda_0 \phi
\left( \Psi_{1a}^{\dagger} M_1 \Psi_{1a} + \Psi_{2a}^{\dagger} M_2
\Psi_{2a} \right) \Big],
\label{dsid4}
\end{equation}
where $\lambda_0$ is the required linear coupling constant between the order
parameter and a fermion bilinear. The matrices $M_1$, $M_2$ are
given by
\begin{eqnarray}
&& \mbox{(A) $M_1 = \tau^y$, $M_2 = \tau^y$} \nonumber \\
&& \mbox{(B) $M_1 = \tau^y$, $M_2 = -\tau^y$} \nonumber \\
&& \mbox{(C) $\lambda_0 = 0$} \nonumber \\
&& \mbox{(D) $M_1 = \tau^x$, $M_2 = \tau^x$} \nonumber \\
&& \mbox{(E) $M_1 = \tau^z$, $M_2 = -\tau^z$} \nonumber \\
&& \mbox{(F) $M_1 = \tau^x$, $M_2 = -\tau^x$}
\label{m1m2}
\end{eqnarray}
Note that there is no non-derivative coupling between $\phi$ and
$\Psi$ for case C: the fermions are essentially spectators of
the transition for this case, which will not be considered
further. Finally, for case G, $S_{\Psi\phi}$ generalizes to
\begin{equation}
\widetilde{S}_{\Psi\phi} = \int \!\! d^2 x d \tau \Big[ \lambda_0
\left(\phi_x \Psi_1^{\dagger} \Psi_1 + \phi_y \Psi_2^{\dagger}
\Psi_2 \right) \Big].
\label{p2}
\end{equation}

We can also consider the coupling between the order parameter $\phi$
and the phase of the superconducting order, $\varphi$, in
(\ref{orders}). By symmetry, the simplest allowed coupling
is $\phi^2 (\nabla \varphi)^2$. It is easy to show that such a
coupling is irrelevant. So the power-law correlations generated by
the superflow are not important.

We now describe our RG analysis of the 6 distinct field theories
represented by $S_{\Psi} + S_{\phi} + S_{\Psi\phi}$ and
$S_{\Psi} + \widetilde{S}_{\phi} + \widetilde{S}_{\Psi\phi}$.
The familiar momentum-shell method, in which degrees of freedom with
momenta between $\Lambda$ and $\Lambda-d\Lambda$ are successively
integrated out, fails:
the combination of momentum dependent
renormalizations at one loop, the direction-dependent
velocities ($v_F$, $v_{\Delta}$, $c$ \ldots), and the hard
momentum cut-off generate unphysical non-analytic terms in the
effective action. So we will construct RG equations
by~\cite{bgz} using a soft cut-off at scale $\Lambda$, and by
taking a $\Lambda (d / d \Lambda)$ derivative of the renormalized vertices
and self energies.
We write the bare propagators as
\begin{equation}
G_{\Psi_1} = \frac{i p_1 + v_F p_2 \tau^z + v_{\Delta} p_3 \tau^x}{p_1^2
+ v_F^2 p_2^2 + v_{\Delta}^2 p_3^2}K(p^2/\Lambda^2)
\end{equation}
and
\begin{equation}
G_\phi = \frac{1}{p^2}K(p^2/\Lambda^2)
\end{equation}
where $K(y)$ is some decaying cuf-off function with $K(0)=1$; {\em e.g.},
$K(y) = e^{-y}$ is a convenient choice.
[The momentum-shell method would correspond to a step-function
cut-off $K(y) = \Theta(y-1)$.]
To handle possible anisotropies we use a hybrid approach;
the space-time integrals over ${\bf p} = (\omega_n,p_x, p_y)$
are written down in $D=2+1$ dimensions,
they can be split via ${\bf p} = p\, {\bf n}$
(where $\bf n$ is a unit vector) into an angular integral
$\int d \Omega_n$ containing all direction
dependent information and an integral over $p = |{\bf p}|$ which
is then evaluated in $D=4-\epsilon$ dimensions.

\begin{figure}[t]
\centering \includegraphics[width=5in]{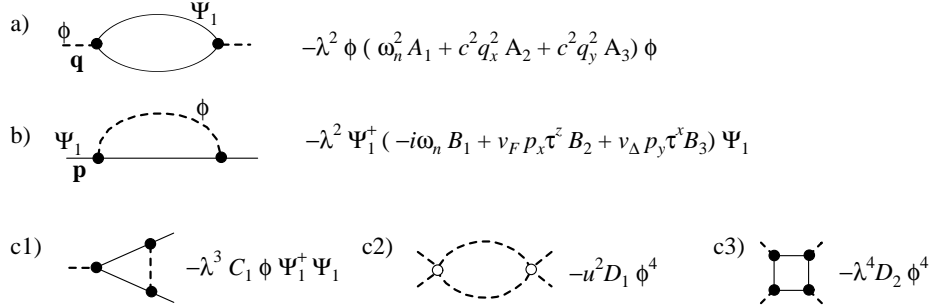} 
\caption{
Diagrams entering the one-loop RG equations for cases A,B,D--F.
Full lines (dashed) lines denote fermionic (bosonic) propagators,
filled (open) circles represent the $\lambda_0$ ($u_0$) interaction.
The constants $A_i$, ... are obtained by evaluating the diagrams, expanding
to lowest non-trivial order in the external momentum and taking
the $\Lambda (d / d \Lambda)$ derivative as described in the text.
a) and b) are the velocity renormalizations whereas c) contains the
renormalized couplings.
(Diagrams for $\Psi_2$ are similar.)
}
\label{figdgr}
\end{figure}

We demonstrate this method by calculating the linear-order
fermionic self-energy $\Sigma_{\Psi_1}$ for the cases A,B,D--F.
The diagram shown in Fig~\ref{figdgr}b evaluates to
\begin{eqnarray}
\Sigma_{\Psi_1}({\bf q}) =
\lambda_0^2 \int_{0}^{\infty} \! \frac{p^2 dp}{8\pi^3}
\int \! d \Omega_n
K\!\!\left(\frac{p^2}{\Lambda^2}\right)
K\!\!\left(\frac{(\bf{p}\!-\!\bf{q})^2}{\Lambda^2}\right)
% \nonumber \\ \times
\frac{i p_1 + s_2 v_F p_2 \tau^z + s_3 v_{\Delta} p_3 \tau^x}
{(p_1^2 + v_F^2 p_2^2 + v_{\Delta}^2 p_3^2)(\bf{p}-\bf{q})^2}
\,.\nonumber
\end{eqnarray}
For $\Sigma_{\Psi_1}$ each vertex contains the coupling matrix $M_1$,
the signs $s_2$, $s_3=\pm 1$ are therefore given by
$M_1 \tau^z M_1 = s_2 \tau^z$, $M_1 \tau^x M_1 = s_3 \tau^x$.
Expanding the above expression to linear order in ${\bf q}$ gives:
\begin{eqnarray}
\Sigma_{\Psi_1}(\bf{q})
&=&
2 \lambda_0^2 \int_{\mu}^{\infty} \frac{ dp}{8 \pi^3}
\left[ \frac{-p^2 K(p^2/\Lambda^2)
K^{\prime} (p^2/\Lambda^2) + \Lambda^2 K^2 (p^2/\Lambda^2)}{p^2 \Lambda^2} \right]
\nonumber \\
&&~~~~~~~~~\times \int d \Omega_n
\frac{i n_1^2 q_1 + s_2 v_F n_2^2 q_2 \tau^z + s_3 v_{\Delta} n_3^2 q_3 \tau^x}
{n_1^2 + v_F^2 n_2^2 + v_{\Delta}^2 n_3^2}
\,.\label{e1}
\end{eqnarray}
We have inserted a lower limit $\mu$ in the $p$ integral
representing an external momentum to regularized the
infrared divergence.
Note that the only quantity entering the RG equations is
the $\Lambda (d / d \Lambda)$ derivative of the self-energy.
This removes the infrared divergence, {\em i.e.}, we can take the limit
$\Lambda/\mu \rightarrow \infty$ in the final expression.
Writing the cut-off integral involving $K$ in general dimension $D$
we obtain
\begin{eqnarray}
\Lambda \frac{d}{d \Lambda} \Sigma_{\Psi_1} &=&
 \frac{2 \lambda_0^2 K_1}{(2 \pi)^D \Lambda^{4-D}}
 \int d \Omega_n \frac{i n_1^2 q_1 + s_2 v_F n_2^2 q_2 \tau^z
 + s_3 v_{\Delta} n_3^2 q_3 \tau^x}
 {n_1^2 + v_F^2 n_2^2 + v_{\Delta}^2 n_3^2}
\,, \label{e2}  \\
K_1 &=& \int_{0}^{\infty} d y\, y ^{(D-4)/2}
\left[ - K(y)K^{\prime} (y) + y [K^{\prime} (y)]^2 + y
K(y) K^{\prime\prime} (y) \right]
\,. \nonumber
\end{eqnarray}
A simple integration by parts for $D=4$ shows that $K_1$
equals unity, {\em independent} of $K(y)$.

At this point it is useful to introduce dimensionless coupling constants $\lambda$ and $u$ by
$\lambda_0 = \lambda \, \Lambda^{\epsilon/2}/S_D^{1/2}$ and
$u_0 = u \, \Lambda^\epsilon/S_D$ with
$S_D = \Omega_D/(2\pi)^D = 2/[\Gamma(D/2)(4\pi)^{D/2}]$ and
$\Omega_D$ being the area of a $D$-dimensional unit sphere.
From Eq~(\ref{e2}) we can easily read off the values of the $B_i$:
\begin{eqnarray}
B_i = - 2 K_1 \int \frac{d \Omega_n}{\Omega_D} \,\frac{s_i n_i^2}{n_1^2 + v_F^2 n_2^2 + v_{\Delta}^2 n_3^2}
\label{bi}
\end{eqnarray}
with $s_1=-1$.
% and we have absorbed the constant prefactors in (\ref{e2}) into the definition of $B_i$.
The other diagrams in Fig~\ref{figdgr} are evaluated similarly.

The derivation of the RG equations is straightforward and essentially identical
to the momentum shell method.
We construct flow equations for all velocities and couplings to
one-loop order in the non-linearities $\lambda$, $u$.
It is convenient to choose the time scale such that $c=1$ at each step
(this introduces a non-trivial dynamical critical exponent $z$).
The procedure results in the following RG equations for cases A,B,D--F:
\begin{eqnarray}
\beta(v_F)      &=& - v_F \lambda^2 \left ( B_2 - B_1 + \frac{\bar{A_1}-\bar{A_2}}{2} \right ) \,, \nonumber \\
\beta(v_\Delta) &=& - v_\Delta \lambda^2 \left ( B_3 - B_1 + \frac{\bar{A_1}-\bar{A_2}}{2} \right ) \,, \nonumber \\
\beta(\lambda)  &=& - \frac{\epsilon}{2} \lambda + \lambda^3
                      \left ( \frac{3 \bar{A_2} - \bar{A_1} }{4} + B_1-C_1 \right ) \,, \nonumber \\
\beta(u)        &=& - \epsilon u + \frac{\lambda^2 u}{2} (\bar{A_1} + 3 \bar{A_2}) - \frac{u^2}{24} D_1 - 24 \lambda^4 D_2
\label{betaf}
\end{eqnarray}
with the renormalization constants as defined in Fig~\ref{figdgr} and $\bar{A_1} = 2 A_1$,
$\bar{A_2} = A_2+A_3$.
At a fixed point the dynamical critical exponent $z$ and the anomalous field dimensions are given by
\begin{eqnarray}
z &=& 1 + \lambda^{\ast 2} (\bar{A_1}-\bar{A_2})/ 2 \,, \nonumber \\
\eta_b &=& \lambda^{\ast 2} \bar{A_2} \,, \nonumber \\
\eta_f &=& \lambda^{\ast 2} \left[ B_1 - (\bar{A_1}-\bar{A_2}) / 2 \right] \,.
\end{eqnarray}

From Eq~(\ref{betaf}) we see that $\beta(v_F)=\beta(v_\Delta)=0$ requires $B_2=B_3$
at a non-trivial fixed point.
The signs of the $B_i$ (\ref{bi}) are determined by $s_i$, and this preempts
a fixed point for the cases D--F: the structure of $M_{1,2}$ leads to
$s_2 = -s_3$, so that $B_2$ and $B_3$ have always different signs.

The cases A and B have $s_2=s_3=-1$, and it is easy to show that the fixed point equations
for $v_F$ and $v_\Delta$ are simultaneously satisfied only for
$v_F^\ast=v_\Delta^\ast=1$.
This implies that the resulting fixed point is Lorentz invariant~\cite{vzs};
for $D=4$ the renormalization constants are given by
$\bar{A}_{1,2}=4$, $B_i=1/2$, $C_1=-1$, $D_1=-36$, $D_2=2$.
The RG equations have the infrared stable fixed point~\cite{vzs} with $z=1$ and
\begin{equation}
\lambda^{\ast 2} = \frac{\epsilon}{7}\,,~~
u^\ast = \frac{16\epsilon}{21}\,.
\end{equation}

Let us briefly discuss the remaining case G.
By choosing the time scale we keep the velocity $c_1 = 1$ fixed.
The flow equations for $v_F$, $v_\Delta$, and $\lambda$ have the same form
as above, but with $\bar{A}_{1,2}$ replaced by $A_{1,2}$, since each fermion field
couples to only one species of bosons.
The other RG equations read:
\begin{eqnarray}
\beta(e)        &=& e \lambda^2 A_2\,,~~~~~~~
\beta(c_2^2)     =  c_2^2 \lambda^2 (A_3 - A_2)\,, \nonumber \\
\beta(u)        &=& - \epsilon u + \frac{\lambda^2 u}{2} (A_1 + 3 A_2)
                    - \frac{u^2 + v^2}{24} D_1 - 24 \lambda^4 D_2 \,,  \nonumber \\
\beta(v)        &=& - \epsilon v + \frac{\lambda^2 v}{2} (A_1 + 3 A_2) - \frac{u v}{24} E_1 - \frac{v^2}{24} E_2
\end{eqnarray}
where $E_1$ and $E_2$ renormalizing $v$ arise from diagrams similar to
Fig~\ref{figdgr}c2.
Analysis of these equations shows that $\beta(v_F)=\beta(v_\Delta)=\beta(c_2^2)=0$ is only
fulfilled for $c_2=1$ and $v_F=v_\Delta$.
Furthermore we must have $e=0$ at a fixed point, stability requires $A_2>0$.
Explicit evaluation in $D=4$ shows that the zeros of $\beta(v_F)$
have $A_2<0$ which proves the non-existence of a fixed point for case G.

We summarize this subsection by restating the main conclusions.
Among the ${\bf Q}=0$ order parameters considered here, only cases
A and B possess stable RG fixed points which can describe a
second-order quantum phase transition to the state $X$, and with
strong damping of quasiparticle spectral functions. The fixed
points for both cases are Lorentz invariant, and for future
convenience, we display this explicitly. We perform the unitary
transformation
\begin{equation}
\Psi_{2a} = \frac{(\tau^x + \tau^z)}{\sqrt{2}} \Psi_{\dot{2}a}
\end{equation}
(this ensures that the structure of the $\tau$ matrices in
(\ref{dsid1}) is the same for $\Psi_1$ and $\Psi_{\dot{2}}$),
define $\overline{\Psi}_{1,\dot{2}a} = - i \tau^y \Psi_{1,\dot{2}a}^{\dagger} $,
and introduce the Lorentz index $\mu = \tau,x,y$.
Then $S_{\Psi} + S_{\phi} + S_{\Psi\phi}$ can be written as
\begin{eqnarray}
S_{1} &=& \int d^3 x \Big( i \overline{\Psi}_{1a}
\partial_{\mu} \gamma^{\mu} \Psi_{1a} + i \overline{\Psi}_{\dot{2}a}
\partial_{\mu} \gamma^{\mu} \Psi_{\dot{2}a} + \frac{1}{2}( \partial_{\mu}
\phi)^2  + \frac{r}{2} \phi^2
 + \frac{u_0}{24} \phi^4 \nonumber \\
 &~&~~~~~~~~~~~~~~~~~~~~~~~~~~~~~~~~~~  - i \lambda_0 \phi (
\overline{\Psi}_{1a} \Psi_{1a} \mp \overline{\Psi}_{\dot{2}a} \Psi_{\dot{2}a}) \Big),
\label{dsid7}
\end{eqnarray}
where $\gamma^{\mu} = (-\tau^y, \tau^x, \tau^z)$, we have
set $v_F=v_{\Delta}=c=1$ and the $-$ ($+$)
sign in the last term is for case A (B).

\subsection{Order parameters with ${\bf Q} = (2K,2K)$}
\label{afm}

One plausible candidate for ordering at such a wavevector is
antiferromagnetic order at ${\bf Q} = (\pi,\pi)$, $K=\pi/2$;
this case was discussed in Section~\ref{intro}, and has been
analyzed by Balents {\em et al}.~\cite{bfn}
(The generalization to magnetic order at incommensurate wavevectors
is not difficult, but we will not consider it because
incommensuration along the diagonal direction in the Brillouin
zone has not been observed in the superconducting cuprates.) We
represent the strength of the antiferromagnetic order by a real,
three-component field $N_{\alpha}$ ($\alpha =x,y,z$; for the
incommensurate case $N_{\alpha}$ is complex). The
transition from a $d$-wave superconductor with $\langle N_{\alpha}
\rangle =0$ to a state $X$ which is a $d$-wave superconductor with
$\langle N_{\alpha} \rangle \neq 0$ is described by the following
continuum action near the critical point~\cite{bfn}
\begin{eqnarray}
S_{2} &=& \int d^3 x \Big( i \overline{\Psi}_{1a}
\partial_{\mu} \gamma^{\mu} \Psi_{1a} + i \overline{\Psi}_{\dot{2}a}
\partial_{\mu} \gamma^{\mu} \Psi_{\dot{2}a} + \frac{1}{2}( \partial_{\mu}
N_{\alpha})^2  + \frac{r}{2} N_{\alpha}^2
 + \frac{u_0}{24} (N_{\alpha}^2)^2 \nonumber \\
 &~&~~~~~~~~~+ i \lambda_0 N_{\alpha} \varepsilon_{ac} \sigma^{\alpha}_{cb} \left[
\Psi_{1a} {\cal C} \Psi_{1b}
- \Psi_{\dot{2}a} {\cal C} \Psi_{\dot{2}b}
+ \mbox{H.c.} \right] \Big),
\label{bfn}
\end{eqnarray}
where $\sigma^{\alpha}$ are Pauli matrices in spin space, and
${\cal C} = i \tau^{y}$ is a matrix in Nambu space.
Like $S_1$, $S_2$ has the property of Lorentz invariance
($\Psi {\cal C} \Psi$ and $\overline{\Psi} \Psi$ are Lorentz
scalars). Also like $S_1$, the couplings $\lambda$ and $u$
approach~\cite{bfn} fixed point values under the RG, and so the
fermion spectral function in the quantum-critical region will have
an energy width of order $k_B T$.

For completeness, we mention another order parameter at ${\bf
Q}=(\pi,\pi)$ which has been the focus of some recent discussion:
the staggered flux order~\cite{flux}. The coupling of this order to the nodal
fermions involves a spatial derivative, and can be shown to be
irrelevant:~\cite{vzs} so the fermionic quasiparticles are merely
spectators at a transition involving the onset of staggered flux
order and such a theory is not of interest here.

\subsection{Order parameter with ${\bf Q} = (2K,0), (0, 2K)$}
\label{cdw}

An attractive possibility of an order parameter with such a ${\bf
Q}$ is the charge density itself (or `charge stripe' order).
We consider the incommensurate case, in which the charge density
has the modulation
\begin{equation}
\delta\rho = \mbox{Re} \left[ \Phi_x e^{i 2K x} + \Phi_y e^{i 2K y}
\right]
\label{stripe}
\end{equation}
where $\Phi_{x,y}$ are complex fields which
constitute the charge-stripe order parameter. Related order
parameters have been considered earlier by Castellani {\em et al}
\cite{castro}, but they discussed the onset of charge density wave
order in a Fermi liquid (for this case $K$ has no
meaning, and one considers instead charge density wave order at a generic
${\bf Q}$ which can connect two points on the Fermi surface),
whereas we will consider the onset in a $d$-wave superconductor.
The theories for the two cases are very different: the
former has an overdamped spectrum for the order parameter
fluctuations \cite{hertz} and (because the theory is not below
its upper-critical dimension) does {\em not} satisfy the strong
scaling properties assumed for the fermion spectral functions to
be discussed in Section~\ref{single}, which do apply to the latter.
We believe it is
important to discuss the onset of charge order in the true ground
state of the doped antiferromagnet, the $d$-wave superconductor,
and hence our focus on this case.

The field theory for
a transition from a $d$-wave superconductor to a state $X$ which
is a $d$-wave superconductor co-existing with the modulation
(\ref{stripe}) can be deduced~\cite{vs,vzs} as in the cases above,
and near the critical fixed point it takes the form
\begin{eqnarray}
S_{3} &=& \int d^3 x \Big( i \overline{\Psi}_{1a}
\partial_{\mu} \gamma^{\mu} \Psi_{1a} + i \overline{\Psi}_{\dot{2}a}
\partial_{\mu} \gamma^{\mu} \Psi_{\dot{2}a} + |\partial_{\mu}
\Phi_x|^2 +  |\partial_{\mu}
\Phi_y|^2+ r \left( |\Phi_x|^2 + |\Phi_y|^2 \right)
 \nonumber \\ &~& \!\!\!\!\!\!\!\!\!\!\!
+ \frac{u_0}{2} \left( |\Phi_x|^4 \!+\! |\Phi_y|^4 \right)
+ v_0 |\Phi_x|^2 |\Phi_y|^2
+ \lambda_0 \! \left[
\Phi_x \overline{\Psi}_{\dot{2}a}  \Psi_{1a}
\!-\! \Phi_y \varepsilon_{ab} \Psi_{\dot{2}a} {\cal C} \Psi_{1b}
\!+\! \mbox{H.c.} \right] \Big).
\label{s3}
\end{eqnarray}
The RG properties~\cite{vzs} of $S_3$ are very similar to the cases
$S_{1,2}$ considered previously: $u$, $v$, and $\lambda$ approach
fixed point values, and there is a universal coupling between
fermionic and bosonic fluctuations in the critical region.

\section{Fermion spectral functions}
\label{single}

The remainder of this paper will highlight some important
observable properties of the field theories $S_{1,2,3}$ in
(\ref{dsid7},\ref{bfn},\ref{s3}). We will restrict ourselves to
the scaling limit, both at $T=0$ and $T>0$; for some quantities we
will see that it is necessary to consider corrections to scaling
to obtain a complete picture---we will not enter into such issues
here and leave them for future work.

In this section we consider single fermion Green's functions which
can be measured in photoemission experiments. In the
quantum-critical region (Fig ~\ref{fig1}) of $S_{1,2,3}$ these
will obey
\begin{equation}
G_f ( k , \omega) = \frac{{\cal A}_f}{T^{(1-\eta_f)}} \,
\Phi_f \!\left( \frac{\omega}{T}, \frac{k}{T} \right),
\label{scal}
\end{equation}
where we have set $\hbar$, $k_B$ and all velocities to unity. The
scale factor ${\cal A}_f$ is non-universal, while the exponent
$\eta_f$ and the complex-valued function $\Phi_f$ are universal
and depend only upon whether the relevant theory is $S_1$, $S_2$
or $S_3$. Scaling functions like $\Phi_f$ have been studied by a
variety of approximate methods~\cite{book}, and we will quote
some useful limiting forms here (in contrast, in 1+1 dimensions exact results
for analogous scaling functions are known~\cite{book,orgad}
for all values of their arguments).

For $\omega,k \gg T$, (\ref{scal}) reduces to \cite{bfn,vzs}
\begin{equation}
G_f (k, \omega) =
{\cal A} C_f \frac{ \omega + k_x \tau^z + k_y
\tau^x}{\left[ k^2 - (\omega+ i0)^2 \right]^{1-\eta_f/2}},
\label{tail}
\end{equation}
where $C_f$ is a universal number.
Note that the imaginary part of this is non-zero only for $\omega
> k$, and it decays as $\omega^{-1+\eta_f}$ for large $\omega$; a
similar kinematic constraint and large $\omega$ tail was noted
recently in one-dimensional stripe models for fermion spectral
functions \cite{orgad}. The expression (\ref{tail}) has a
singularity precisely at $\omega=k$: this singularity is rounded
at any non-zero $T>0$ and so~(\ref{tail}) is not an accurate
representation of (\ref{scal}) right at the threshold.
Unfortunately, the nature of this rounding is not easy to compute
in general; it can be computed in an expansion in $\eta_f$, and
these results are in Appendix A of our recent work~\cite{vzs}.

In the opposite limit, $\omega,k \ll T$, (\ref{scal}) has a very
different form. Now the $k$ and $\omega$ dependencies are smooth,
and all anomalous powers involve only factors of $T$. The
following approximate form was suggested \cite{vzs} in this limit
\begin{equation}
G_{f} (k, \omega) = {\cal A} T^{\eta_f}
\frac{\omega + i \Gamma_f + k_x \tau^z + k_y \tau^x}{k^2 - (\omega
+ i \Gamma_f)^2 },
\label{damp}
\end{equation}
where the overall scale of the right hand side defines the value
of ${\cal A}$, and $\Gamma_f /T$ is a universal number. Estimates
of this universal number have been given for $S_{1,3}$. Note that
along the diagonal direction in the Brillouin zone ($k_x=0$),
the result (\ref{damp}) predicts a simple Lorentzian for the
spectral function with an energy width of order $\Gamma_f$.
Of course, the Lorentzian form breaks down for large $\omega$,
when the spectral function crosses over to the slowly decaying
tail predicted by (\ref{tail}).

\section{Transport properties}
\label{trans}

Unlike the single-particle properties of Section~\ref{single},
transport properties can require determination of composite
correlators of both the fermionic and bosonic degrees of freedom.
In particular, we can have processes in which there is rapid
scattering between fermions and bosons, and thus a broad fermion
spectral function, but essentially no degradation of the transport
current. In such cases, it is clearly not appropriate to interpret
transport properties in terms of a single fermionic quasiparticle
lifetime, and attempts to do so will suggest very long
quasiparticle lifetimes very different from those observed in
photoemission experiments.

We describe spin, thermal, and charge transport in
the following subsections:

\subsection{Spin transport}
\label{spin}

The global $SU(2)$ spin-rotation symmetry of $S_{1,2,3}$ implies
that total spin is conserved. Consequently, there is an associated
Lorentz 3-current, $J^{s \alpha}_{\mu}$ obeying $\partial_{\mu}
J^{s \alpha}_{\mu} = 0$. For $S_{1,3}$ this 3-current has only a
fermionic contribution
\begin{equation}
J^{s \alpha}_{\mu} = \frac{1}{2} \left[
\overline{\Psi}_{1a} \sigma^{\alpha}_{ab} \gamma_{\mu} \Psi_{1b}
+ \overline{\Psi}_{\dot{2}a} \sigma^{\alpha}_{ab} \gamma_{\mu}
\Psi_{\dot{2}b} \right]
\label{st1}
\end{equation}
(this expression agrees with earlier work~\cite{durst}),
while for $S_2$ there is also bosonic contribution \cite{book}
\begin{equation}
J^{s \alpha}_{\mu} = \frac{1}{2} \left[
\overline{\Psi}_{1a} \sigma^{\alpha}_{ab} \gamma_{\mu} \Psi_{1b}
+ \overline{\Psi}_{\dot{2}a} \sigma^{\alpha}_{ab} \gamma_{\mu}
\Psi_{\dot{2}b}\right] + i \epsilon_{\alpha\beta\gamma} N_{\beta}
\partial_{\mu} N_{\gamma}.
\label{st2}
\end{equation}
The conservation of $J^{s \alpha}_{\mu}$ implies that the total
spin is a constant of motion
\begin{equation}
\frac{d}{d \tau} \int d^2 x J^{s \alpha}_{\tau} = 0;
\label{st3}
\end{equation}
however, the physical spin current is not conserved:
\begin{equation}
\frac{d}{d \tau} \int d^2 x J^{s \alpha}_{x,y} \neq 0.
\label{st4}
\end{equation}
The spin conductivity, $\sigma_s$, is expressed by a Kubo formula
involving $J^{s \alpha}_{x,y}$ (along with a `diamagnetic' contact
term for the bosonic fields in $S_2$) and (\ref{st4}) implies that
a ballistic contribution to $\sigma_s$ is not expected. The
scaling properties of $\sigma_s$ can be deduced along the lines
discussed \cite{book} for a purely bosonic theory: the
conservation of the 3-current implies that the scaling dimension
of $\sigma_s$ is protected to take the value $d-2=0$, and so in
the quantum critical region we have~\cite{book,science}
\begin{equation}
\sigma_s = \Phi_s \left( \frac{\omega}{T} \right),
\end{equation}
where $\Phi_s$ is a universal function of order unity (we have
absorbed factors of $g\mu_B$ in the definition of $\sigma_s$).
In principle, it should not be difficult to extend earlier
results~\cite{book} for purely bosonic theories to determine the
collisionless ($\omega \gg T$) and collision-dominated ($\omega \ll
T$) regimes of $\sigma_s$.

\subsection{Thermal transport}
\label{heat}

This section is based on ideas of T.~Senthil (unpublished) in a
different context.

The Lorentz invariance of $S_{1,2,3}$ implies the existence of a
energy-momentum 3-tensor $\Theta_{\mu\nu}$ which is
conserved, $\partial_{\mu}\Theta_{\mu\nu} =0$, and which can be chosen to symmetric
in the $\mu,\nu$ indices \cite{wein}.
The energy density is
$\Theta_{\tau\tau}$, while the energy current is $J^{e}_i =
\Theta_{i\tau}$, where $i=x,y$. The conservation of the 3-tensor
$\Theta_{\mu\nu}$ implies three constants of motion,
the total energy and the total momentum:
\begin{equation}
\frac{d}{d\tau} \int d^2 x \Theta_{\tau\nu} = 0.
\end{equation}
Now using the symmetry of $\Theta_{\mu\nu}$, and in particular
$\Theta_{i\tau} = \Theta_{\tau i}$, we conclude that the spatial integral of
the energy current
$J^{e}_i$ is also conserved [contrast this with (\ref{st4}), in
which the spatial integral over the spin current was not
conserved].
As the thermal conductivity is given by a Kubo formula involving
the energy current, we conclude that heat transport is ballistic
and the thermal conductivity is infinite in the scaling limits
defined by $S_{1,2,3}$.

\subsection{Charge transport}
\label{elec}

The above theories describe excitations in superconductors,
and a complete description of
charge transport, and an accounting of the conservation of
total charge, requires inclusion of the Cooper
pairs [alternatively stated, we have to include the fluctuations
of $\varphi$ in (\ref{orders})]. However, in
a frequency regime where the response is dominated by
quasiparticle excitations,
we can consider the predictions of $S_{1,2,3}$ alone. The quasiparticle
contribution to the electrical current is~\cite{durst}
the 2-vector $J^{c}_i =
(\Psi_{1a}^{\dagger} \Psi_{1a} , \Psi_{2a}^{\dagger} \Psi_{2a})
= i (\overline{\Psi}_1 \gamma_{\tau} \Psi_1,
-\overline{\Psi}_{\dot{2}a} \gamma_{\tau} \Psi_{\dot{2}a})$.
In general there is nothing special about this operator: it will
acquire anomalous dimensions at the critical point, and this will
lead to a contribution to the charge conductivity which has
prefactors of a non-universal constant and an anomalous power of
$T$. However, for the case of $S_1$ only, an exceptional
circumstance occurs: the action has $U(1)$ symmetries
corresponding to global changes in the phases of
$\Psi_{1,\dot{2}}$, and corresponding conserved 3-currents
$\partial_{\mu} \overline{\Psi}_{1a} \gamma_{\mu} \Psi_{1a}
= \partial_{\mu} \overline{\Psi}_{\dot{2}a} \gamma_{\mu}
\Psi_{\dot{2}a}=0$. Note that these 3-currents do {\em not}
correspond to the conservation of physical electrical charge:
rather, the $\tau$ component of these 3-currents happen
to be equal to the $x,y$ components of the electrical current
$J^{c}_i$. The conservations of these currents implies that the
spatial integral of $J^{c}_i$ is a constant of motion, and
hence the quasiparticle contribution to the charge conductivity is
a zero-frequency delta function for $S_1$.

\section{Conclusion}

This paper has identified quantum field theories at which the
nodal quasiparticles of a $d$-wave superconductor undergo strong
inelastic scattering. All these theories are associated with a
quantum critical point between a $d$-wave superconductor and some
other superconducting state $X$ (Fig~\ref{fig1}). We performed an
exhaustive search among states $X$ associated with a spin-singlet,
zero momentum, fermion bilinear order parameter, and found that
only two candidates possessed a non-trivial quantum
critical point: $X$ a $d_{x^2-y^2}+is$ or a
$d_{x^2-y^2}+id_{xy}$ superconductor. The quantum field theory for
these cases is $S_1$ in (\ref{dsid7}). Among order
parameters with a non-zero momentum, we had to permit a
fine-tuning condition that the ordering momentum was equal to the
spacing between two Fermi points of the $d$-wave superconductor.
Two plausible cases of this kind had $X$ possessing co-existing
spin-density wave order and $d$-wave superonductivity [described by
$S_2$ in (\ref{bfn})] or co-existing charge-density wave order
and $d$-wave superconductivity [described by $S_3$ in (\ref{s3})].

We reviewed the single particle (Section~\ref{single}) and
transport (Section~\ref{trans}) properties of $S_{1,2,3}$.
An important observation was that there is no simple relationship
between the relaxation times associated with these observables. In
particular, strong scattering between the fermionic and bosonic
modes can lead to very short single particle lifetimes and broad
spectral functions (Section~\ref{single}). At the same
time, these scattering processes can do little to degrade the
transport current: the thermal conductivity of $S_{1,2,3}$ was
found to be infinite in the continuum scaling limit.

Comparison of the above results with experiments requires some
understanding of the effects of corrections to scaling, and of
impurity scattering. Initial steps in this direction
have been taken~\cite{vzs}, and studies for transport
properties are in progress.

Keeping the above cautions in mind, we offer a possible
interpretation of
recent thermal conductivity measurements \cite{ong}. A strong
enhancement of the thermal Hall conductivity, $\kappa_{xy}$, is observed as $T$
is lowered, but $\kappa_{xy}$ saturates and decreases below $T=28$
K. The enhancement suggests large fluctuations of chiral order\cite{chandra},
as is the case when $X$ is a $d_{x^2-y^2} + id_{xy}$
superconductor (for $T < T_X$ this state has a non-zero
$\kappa_{xy}$ even in zero field \cite{senthil}). So we suggest
that the experimental system is in the vicinity of such a state
$X$ but has $r>r_c$, and the crossover at the dashed line in
Fig~\ref{fig1} occurs at $T \approx 28$ K.
A testable implication is that
the photoemission linewidths of the nodal fermions should be $\sim
k_B T$ down to 28 K, and then cross over to behavior expected in an
ordinary BCS $d_{x^2-y^2}$ superconductor at lower $T$.

%%%%%%%%%%%%%%%%%%%%%%%%%%%%%%%%%%%%%%%%%%%%%%%%%%%%%%%%%%%%%%

\section*{Acknowledgments}
We thank E.~Carlson, P.~Johnson, N.~P.~Ong, and T.~Senthil for
useful discussions and the US NSF (DMR 96--23181) and the DFG (VO 794/1-1) for support.

%%%%%%%%%%%%%%%%%%%%%%%%%%%%%%%%%%%%%%%%%%%%%%%%%%%%%%%%%%%%%%

\newpage

\title{Erratum: Quantum phase transitions in $d$-wave superconductors\\
Phys. Rev. Lett. {\bf 85}, 4940  (2000)}

\author{Matthias Vojta}
\address{Institut f\"ur Theoretische Physik,
Universit\"at zu K\"oln, Z\"ulpicher Str. 77, 50937 K\"oln, Germany}

\author{Ying Zhang}
\address{Goldman Sachs, 1 New York Plaza, New York, NY 10004}

\author{Subir Sachdev}
\address{Department of Physics, Harvard University, Cambridge,
MA 02138}

\date{\today}

\begin{abstract}
We correct an error in our paper Phys. Rev. Lett. {\bf 85}, 4940 (2000)
[arXiv:cond-mat/0007170].
Our characterization of the physical properties of the
superconducting state G was incorrect:
it breaks time-reversal symmetry, carries spontaneous currents,
and possesses Fermi surface pockets.
\end{abstract}

\bodymatter
~\\

In the discussion of case G in our paper\cite{prl}, above Eq.~(4), the single 
sentence
``The state $X$ retains $\mathcal{T}$ and the gapless nodal
points, but has $C_{4v}$ broken to $Z_2$'' is incorrect.
The state $X={\rm G}$ breaks $\mathcal{T}$ (time-reversal), and has spontaneous
electrical currents. For $\phi_x \neq 0$ and $\phi_y = 0$ (or vice versa) the 
currents
have the same symmetry as those in the state $\Theta_{\rm II}$
discussed by Simon and Varma \cite{varma2}.
Also, as pointed out by Berg {\em et al.} \cite{berg}, the nodal
quasiparticles do not survive in the superconducting state G, but turn into 
Fermi pockets.
The latter conclusion can be verified from the fermion spectrum
obtained by diagonalizing Eqs. (1)+(5) for constant $\phi_{x,y}$.

All other sentences and the conclusions in the paper \cite{prl} remain 
unchanged.

Also, in the companion paper\cite{long}, the only error is in
the sketch of the
fermion excitations in Fig.~2 for case G.

We thank Erez Berg and Cenke Xu for helpful discussions.
This research was supported by NSF grant DMR-0537077
and DFG SFB 608.


\vspace*{-6pt}

\section*{References}
\begin{thebibliography}{99}

\bibitem[*]{addr} New permanent address: Theoretische Physik III, Elektronische Korrelationen
   und Magnetismus, Universit\"{a}t Augsburg, D-86135 Augsburg, Germany.

\bibitem{valla} T.~Valla, A.~V.~Fedorov, P.~D.~Johnson,
B.~O.~Wells, S.~L.~Hulbert, Q.~Li, G.~D.~Gu, and N.~Koshizuka, Science {\bf
285}, 2110 (1999).

\bibitem{ong} Y.~Zhang, N.~P.~Ong, P.~W.~Anderson, D.~A.~Bonn,
R.~Liang, and W.~N.~Hardy, preprint;
K. Krishana, J. M. Harris, and N. P. Ong, Phys. Rev.
Lett. {\bf 75}, 3529 (1995).

\bibitem{vzs} M.~Vojta, Y.~Zhang, and S.~Sachdev, Phys. Rev. B
Sep 1 (2000), cond-mat/0003163.

\bibitem{vzs2} M.~Vojta, Y.~Zhang, and S.~Sachdev, cond-mat/0007170.

\bibitem{CHN} S.~Chakravarty, B.~I.~Halperin, and D.~R.~Nelson,
Phys. Rev. Lett. {\bf 60}, 1057 (1988);
Phys. Rev. B {\bf 39}, 2344 (1989).

\bibitem{rs} N.~Read and S.~Sachdev, Phys. Rev. Lett.
{\bf 62}, 1694 (1989); {\bf 66}, 1773 (1991);
G.~Murthy and S.~Sachdev, Nucl. Phys. B {\bf 344}, 557 (1990).

\bibitem{CSY} S.~Sachdev and J.~Ye, Phys. Rev. Lett. {\bf
69}, 2411 (1992); A.~V.~Chubukov and S.~Sachdev, {\it ibid.}
{\bf 71}, 169 (1993); A.~V.~Chubukov, S.~Sachdev, and J.~Ye,
Phys. Rev. B {\bf 49}, 11919 (1994).

\bibitem{science} S. Sachdev, Science {\bf
288}, 475 (2000).

\bibitem{nmr} H.~Alloul, P.~Mendels, G.~Collin, and P.~Monod,
Phys. Rev. Lett. {\bf 61}, 746 (1988);
M.~Takigawa, P.~C.~Hammel, R.~H.~Heffner, Z.~Fisk, J.~L.~Smith,
and R.~B.~Schwarz, Phys. Rev. B {\bf 39}, 300 (1989);
T.~Imai, C.~P.~Slichter, K~Yoshimura,
and K.~Kosuge, Phys. Rev. Lett. {\bf 70}, 1002 (1993);
T.~Imai, C.~P.~Slichter, K.~Yoshimura,
M.~Katoh, K.~Kosuge, {\it ibid.} {\bf 71}, 1254 (1993);
A. Sokol and D. Pines, {\it ibid.} {\bf 71}, 2813 (1993);
Y.~Zha, V.~Barzykin, and D.~Pines
Phys. Rev. B {\bf 54}, 7561 (1996);
A.~W.~Hunt, P.~M.~Singer,
K.~R.~Thurber, T.~Imai, {\it ibid.} {\bf 82},
4300 (1999); S.~Fujiyama, M.~Takigawa, Y.~Ueda, T.~Suzuki,
N.~Yamada, Phys. Rev. B {\bf 60}, 9801 (1999).

\bibitem{gabe} G.~Aeppli, T.~E.~Mason, S.~M.~Hayden,
H.~A.~Mook, and J.~Kulda, Science {\bf 278}, 1432 (1997);
S.~M.~Hayden, G.~Aeppli, H.~A.~Mook, D.~Rytz, M.~F.~Hundley, and Z.~Fisk,
Phys. Rev. Lett. {\bf 66}, 821 (1991);
B.~Keimer, N.~Belk, R.~J.~Birgeneau, A.~Cassanho, C.~Y.~Chen, M.~Greven,
M.~A.~Kastner, A.~Aharony, Y.~Endoh, R.~W.~Erwin, and G.~Shirane,
Phys. Rev. B {\bf 46}, 14034 (1992).

\bibitem{pwa} P.~W.~Anderson, cond-mat/0007185.

\bibitem{resonance} J.~Rossat-Mignod, L.~P.~Regnault,
C.~Vettier, P.~Bourges, P.~Burlet, J.~Bossy, J.~Y.~Henry,
and G.~Lapertot,
Physica C {\bf 185-189}, 86 (1991);
H.~A.~Mook, M.~Yehiraj, G.~Aeppli, T.~E.~Mason,
and T.~Armstrong, Phys. Rev. Lett. {\bf 70}, 3490 (1993);
H.~F.~Fong, B.~Keimer, F.~Dogan, and I.~A.~Aksay,
{\it ibid.} {\bf 78}, 713 (1997);
P.~Bourges in {\it The Gap Symmetry and
Fluctuations in High Temperature Superconductors} ed. J.~Bok,
G.~Deutscher, D.~Pavuna, and S.~A.~Wolf, Plenum, New York (1998)
pp. 349-371, cond-mat/9901333;
P.~Dai, H.~A.~Mook, S.~M.~Hayden, G.~Aeppli, T.~G.~Perring,
R.~D.~Hunt, and F.~Dogan {\it Science} {\bf 284}, 1344 (1999);
H.~F.~Fong, P.~Bourges, Y.~Sidis, L.~P.~Regnault, J.~Bossy,
A.~Ivanov, D.~L.~Milius, I.~A.~Aksay, and B.~Keimer, Phys. Rev.
Lett. {\bf 82}, 1939 (1999);
P.~Bourges, Y.~Sidis, H.~F.~Fong, L.~P.~Regnault,
J.~Bossy, A.~Ivanov, and B.~Keimer,
Science {\bf 288}, 1234 (2000);
Y.~Sidis, P.~Bourges, H.~F.~Fong, B.~Keimer, L.~P.~Regnault,
J.~Bossy, A.~Ivanov, B.~Hennion, P.~Gautier-Picard, G.~Collin,
D.~L.~Millius, and I.~A.~Aksay, Phys. Rev. Lett. {\bf 84}, 5900
(2000).

\bibitem{vs} M.~Vojta and S.~Sachdev, Phys. Rev. Lett.
{\bf 83}, 3916 (1999).

\bibitem{bfn} L.~Balents, M.~P.~A.~Fisher, and C.~Nayak, Int. J. Mod.
Phys. B {\bf 12}, 1033 (1998).

\bibitem{caveat} We will not consider the cases where the
separation between nodal points is an integer multiple of ${\bf
Q}$, because the order-parameter/fermion coupling is then found to
be irrelevant by power-counting \protect\cite{vzs}.

\bibitem{bgz} E.~Br\'{e}zin, J.~C.~Le Guillou, J.~Zinn-Justin in
{\em Phase Transitions and Critical Phenomena}, vol {\bf 6}, C.~Domb
and M.~S.~Green eds, Academic Press, London (1976); the method
used by us is discussed in Section 5.

\bibitem{flux} D.~A.~Ivanov, P.~A.~Lee, and X.-G.~Wen,
Phys. Rev. Lett. {\bf 84}, 3958 (2000);
S.~Chakravarty, R.~B.~Laughlin, D.~K.~Morr, and
C.~Nayak, cond-mat/0005443.

\bibitem{castro} C.~Castellani, C.~DiCastro, and M.~Grilli,
Phys. Rev. Lett. {\bf 75}, 4650 (1995);
S.~Caprara, M.~Sulpizi, A.~Bianconi, C.~Di Castro, and
M.~Grilli, Phys. Rev. B {\bf 59},
14980 (1999).

\bibitem{hertz} J.~A.~Hertz, Phys. Rev. B {\bf 14}, 1165 (1976).

\bibitem{book} S.~Sachdev, {\em Quantum Phase Transitions},
Cambridge University Press, Cambridge (1999).

\bibitem{orgad} D.~Orgad, S.~A.~Kivelson, E.~W.~Carlson,
V.~J.~Emery, X.~J.~Zhou, and Z.~X.~Shen, cond-mat/0005457.

\bibitem{durst} A.~C.~Durst and P.~A.~Lee, Phys. Rev. B {\bf 62}, 1270
(2000).

\bibitem{wein} S.~Weinberg, {\em The Quantum Theory of Fields},
vol {\bf 1}, Cambridge University Press, Cambridge (1995), Section 7.4.

\bibitem{chandra} G.~Kotliar, A.~M.~Sengupta, and C.~M.~Varma
Phys. Rev. B {\bf 53}, 3573 (1996).

\bibitem{senthil} T.~Senthil,
J.~B.~Marston, and M.~P.~A.~Fisher, Phys. Rev. B {\bf 60}, 4245
(1999).

\end{thebibliography}

\begin{thebibliography}{99}

\bibitem{prl}
M. Vojta, Y. Zhang, and S. Sachdev,
Phys. Rev. Lett. {\bf 85}, 4940  (2000).
%\bibitem{varma}
%C. M. Varma, Phys. Rev. B {\bf 55}, 14554 (1997);
%Phys. Rev. Lett. {\bf 83}, 3538 (1999).

\bibitem{varma2} M. E. Simon and C. M. Varma, Phys. Rev. Lett. {\bf 89}, 247003 (2002),
see Fig 1b.

\bibitem{berg} E.~Berg, C.-C.~Chen, and S.~A.~Kivelson,
Phys. Rev. Lett. {\bf 100}, 027003 (2008).

\bibitem{long} M.~Vojta, Y.~Zhang, and S.~Sachdev,
Int. J. Mod. Phys. B {\bf 14}, 3719 (2000) [arXiv:cond-mat/0008048].

\end{thebibliography}
\end{document}